# Evaluation of Business-Oriented Performance Metrics in e-Commerce using Web-based Simulation

Ilija S. Hristoski[1] and Pece J. Mitrevski[2]

[1] Faculty of Economics, Marksova St, 133
7500 Prilep, Macedonia
ilija.hristoski@uklo.edu.mk

[2] Faculty of Information and Communication Technologies, Partizanska bb,
7000 Bitola, Macedonia
pece.mitrevski@uklo.edu.mk

**Abstract.** The Web 2.0 paradigm has radically changed the way businesses are run all around the world. Moreover, e-Commerce has overcome in daily shopping activities. For management teams, the assessment, evaluation, and forecasting of online incomes and other business-oriented performance measures have become 'a holy grail', the ultimate question imposing their current and future e-Commerce projects. Within the paper, we describe the development of a Web-based simulation model, suitable for their estimation, taking into account multiple operation profiles and scenarios. Specifically, we put focus on introducing specific classes of e-Customers, as well as the workload characterization of an arbitrary e-Commerce website. On the other hand, we employ and embed the principles of the system thinking approach and the system dynamics into the proposed solution. As a result, a complete simulation model has been developed, available online. The model, which includes numerous adjustable input variables, can be successfully utilized in making 'what-if'-like insights into a plethora of business-oriented performance metrics for an arbitrary e-Commerce website. This project is, also, a great example of the power delivered by InsightMaker®, free-of-charge Web-based software, suitable for a collaborative online development of models following the systems thinking paradigm.

## 1. Introduction

As e-Commerce became a mainstream, more and more companies have gone online, selling their products and/or services worldwide, 24/7. Realizing that the phrase "Without revenue, there is no business!" expresses the core essence of doing business online, e-Commerce companies have focused their efforts, in general, on three crucial aspects: (1) how to attract more e-Customers to their virtual stores, (2) how to convert their visits into revenues, and (3) how to develop a comprehensive, yet realistic online income forecast. Regarding the latter one, it is practically impossible to predict online incomes precisely, especially for new products, new businesses, and/or emerging markets, but it is critically important for e-Commerce companies to create high-quality income estimates. This is because the dynamics of the overall economy trends, the marketplace, and company decision-making can never be captured entirely in advance. This is, partly, also true due to the absence of historical data to which executives, managers, and financial analysts can point to in order to validate their own expectations.

However, building an income forecast from a solid foundation provides a good understanding of an e-Commerce business dynamics. That enables planning marketing operations more effectively and more profoundly. The evaluation of e-Commerce incomes has to consider the environment in which an online business operates, and the inclusion of as much as related 'outer' factors that may affect the ability to motivate and drive e-Customers towards an e-Commerce website, and then complete their transactions, as well. Nonetheless, during recent years, it was also realized that e-Commerce incomes are affected, to a great extent, by numerous 'inner' factors, including psychological, social, cognitive, and emotional factors, all of them portraying the human attitudes and behaviors during online shopping sessions. All of these 'human' factors that are subject to the emerging, interdisciplinary field of behavioral economics, have a great impact on economic decisions of individuals and institutions, and therefore, they have immense consequences for market prices, returns, and resource allocation.

## 2.   E-Commerce Incomes and Other Related Performance Metrics

E-Commerce income (revenue) typically refers to the total amount of money received by the company for goods sold or services provided online during a certain time. Among all other business-oriented metrics (e.g. Revenue per Visit, Revenue per Visitor, Conversion Rate, Average Order Value, Buy-to-Visit Ratio etc.), revenue is the ultimate one that reflects the wealth and current positioning of e-Commerce companies on the global market. Apart from selling goods or services, many online companies generate revenues from multiple, yet different income streams, such as advertising, subscription, transaction fees, or affiliate marketing, altogether known as 'revenue models'. However, the sales revenues remain the keystone of doing business online.

Realizing the importance of estimating e-Commerce sales revenues, many authors suggest almost a standardized way for its calculation [1-6]. They suggest, with negligible differences, that income, $I$ [$], can be assessed by using few business-oriented metrics, including the number of visitors (daily, monthly …), $V$, the conversion rate, $CR$ [%], being a ratio between the number of buyers and visitors, and average order value, $AOV$ [$/order], as in (1). The product of $AOV$ and $CR$ is also known as Revenue per Visitor [$/visitor].

$$I[\$] = AOV[\$/order] \times CR[\%] \times V \qquad (1)$$

In addition, it is also worthy to mention that many websites already offer online calculators for estimating e-Commerce sales incomes [7, 8].

Estimating e-Commerce sales incomes according to (1) is quite straightforward, though somewhat disputable, since it approximates roughly the input variables, which yields a significant estimation error. Furthermore, (1) is purely deterministic by its nature, i.e. it does not include any stochastic parameters. Finally, (1) does not include any behavioral components specific to various e-Customers' classes, nor does it takes into account the workload characterization.

Along with the evaluation of e-Commerce revenues, there is also an increasing trend of assessing a set of performance metrics, suitable for measuring particular e-Commerce website's efficiency in attaining its strategic goals. However, the evaluation of so-called *business-oriented performance metrics* is of a vital interest. The set of these innovative measures, which were introduced in the work of Menascé and Almeida, include the following ones [9-11]:

- *Buy-to-Visit Ratio* [%]; A business-oriented performance metric that measures the average number of sale transactions per visit to a particular e-Commerce website; In other words, it shows how often e-Customers buy each time they visit an e-Commerce website;
- *Revenue Throughput* [$/s]; A business-oriented performance metric showing how much money per time unit (second) have been derived from sales, for a particular e-Commerce website;
- *Potential Loss Throughput* [$/s] or *Lost Revenue Throughput* [$/s]; Another business-oriented performance metric that measures the amount of money in e-Customers' shopping carts that were not converted into sales, because e-Customers have left the website prematurely.

The estimation of the above-mentioned metrics is based on the same premises and principles the evaluation of e-Commerce revenues does. Business-oriented performance metrics are very closely related to the financial effects (revenues, sales, profits, and alike) of a particular e-Commerce website and, therefore, they can be very indicative to the website's management team in the processes of decision-making.

Taking into account previously elaborated shortcomings of the 'standardized' way to evaluate e-Commerce incomes given by (1), and due to the existence of very close linkages between online incomes and business-oriented performance metrics, we propose a significantly different, yet more rigorous approach to estimating e-Commerce sales incomes within this paper. Our approach relies heavily on a Web-based simulation, and the usage of the system dynamics logic. In particular, our aim is to develop both a framework and a methodology, based on a simulation model, which relies entirely on the workload characterization of a hypothetic e-Commerce website and takes into account not only various e-Customer classes but also various operating profiles, i.e. scenarios. The basis for carrying out such evaluation is the model of e-Customer behavior during online shopping sessions.

## 3.  Web-based Simulation

Simulations are crucial for companies in minimizing time, costs, and resource usage needed for accomplishing their business activities, which can lead towards achieving their strategic business goals and gaining a substantial competitive advantage. Utilization of simulation techniques and tools can assist in improving business effectiveness and performances since they can increase their ability to make right decisions in complex and uncertain environments.

The core virtue of simulation is the ability to analyze the behavior of a given system and estimate the outcomes of its operation under various real-time scenarios, for a wide gamut of working parameters and various process variables, even during its design phase, i.e. before it is really built up and exploited. Simulation models are being created as simplifications of real systems to help their designers understand and predict the system's behavior and overcome the inherent complexity and complicated nature of reality. Contrasted to analytical (mathematical) models, simulation models are more comprehensible and more credible, because they require fewer simplifying assumptions and capture more characteristics of the observed system.

The class of computer simulations, which is supported by modern software tools, is based on the development of a computer model using simulation language, package, tool or an integrated environment.

Thanks to the emergence of the Web 2.0 paradigm and open standards, technology has given an opportunity to all companies, including those that deal with e-Commerce, to become more innovative and to gain a substantial competitive advantage. More and more, the Web is being considered an online environment suitable for performing both modeling and simulation tasks. The emerging new innovative and alternative approach to computer simulation, which strives to become *de facto* an adequate replacement of the traditional workstation-based computer simulation, has been named as a 'Web-based simulation' (WBS). It is an integration of the Web with the field of computer simulation, assuming an invocation of computer simulation services over the Web, specifically through a user's Web browser [13-16]. WBS is currently becoming a state-of-the-art discipline, a quickly evolving area of computer science, which is of significant interest for both simulation researchers and simulation practitioners, expected to proliferate and even prevail in the forthcoming years.

## 4.     The System Dynamics Approach and InsightMaker®

System Dynamics (SD) modeling is a powerful method for exploring systems on an aggregate level. By 'aggregate', it is meant that SD models look at collections of objects, not the objects themselves. For instance, an SD model of the e-Customers population would look at the population as a whole, not at the individual e-Customers. If compared to Discrete-Event Simulation (DES), SD uses a quite different approach: it is essentially deterministic by nature and it models a system as a series of stocks and flows, whilst state changes are continuous, resembling a motion of a fluid, flowing through a system of 'reservoirs', connected by 'pipes', whilst the flow can be regulated by 'switches'.

SD models are visually constructed from a set of basic building blocks, known as 'primitives'. However, behind the scene, these primitives are 'converted' into a system of differential equations that describe the modeled system mathematically. Since only the dynamics of extremely small and/or well-known systems could possibly be solved analytically, the dynamics of large and/or ill-known systems requires numerical simulation [19].

The key SD primitives are Stocks, Flows, Variables, and Links.

Stocks are graphically presented by rectangles; they store some kind of 'material', e.g. a population of e-Customers.

Flows, graphically depicted by bolded solid lines with arrows, move the 'material' between stocks; they can be either inflow (input into stocks), or outflow (output from stocks), e.g. a flow of e-Customers' arrival in the online store.

Variables are graphically portrayed by ovals; they can be dynamically calculated values that change over time (governed by an equation) or they can be constants (fixed values), e.g. e-Customer arrival rate.

Links, graphically shown by dashed lines with arrows, show the transfer of information between the different primitives in the model. If two primitives are linked, they are related in some way. Links are generally used in conjunction with variables to build mathematical expressions.

Insight Maker® is an innovative, free-of-charge, Web 2.0-based, multi-user, general-purpose, online modeling and simulation environment, completely implemented in JavaScript, which promotes online sharing and collaborative working. It integrates three general modeling approaches into a unified modeling framework, including: (1) system dynamics, (2) agent-based modeling, and (3) imperative programming [20]. To the best of our knowledge, it is the first, yet the one and only free-of-charge Web 2.0-based Internet service that can deliver a plethora of advanced features to its online users, including Causal Loop Diagrams, Rich Pictures Diagrams, Dialogue Mapping, Mind Mapping, as well as Stock & Flow simulation. All these can offer thorough insights into various aspects of a system's dynamics. By supporting agent-based scenarios, storytelling, and sensitivity analysis, Insight Maker® exhibits a wide gamut of features that not only rival but also, in many cases, outperform the traditional, commercially available simulation software packages.

## 5.   Workload Characterization

The workload of a system can be defined as "the set of all inputs that the system receives from its environment during any given period of time", whilst workload characterization is "the process of precisely describing, in a qualitative and quantitative manner, the global workload of an e-business site" [21]. Since it is difficult to handle real workloads due to a large number of constituting elements, it is more practical to reduce and summarize the information needed to describe the workload. However, the choice of characteristics and parameters that will describe the workload depends solely on the purpose of the study, having minded the fact that the model needs to capture the most relevant characteristics of the real workload. This way, in order to reflect changes in the system and/or in the actual workload, it is possible to gain various insights into the system's behavior simply by changing its model parameters.

We have based the workload characterization of a hypothetic e-Commerce website on three fundamental premises: (1) e-Customers' online shopping behaviors mutually differ, but still, it is possible to classify e-Customers into a finite number of different classes; and (2) e-Customers access the e-Commerce website and invoke the specific e-Commerce functions in a rather unpredictable and stochastic manner [22].

The first premise exhibits the qualitative aspects of workload characterization. Many studies have pointed out the fact that it is possible to distinguish between different classes of e-Customers, regarding their specific online shopping behaviors [23, 24]. Recently, the fields of behavioral economics, buyer psychology, and neuroeconomics have been put in focus due to their great contribution to understanding why and how e-Customers make purchases, which are a proven route to successful marketing, as well as to producing conversions and revenues. By combining research methods from neuroscience, experimental and behavioral economics, psychiatry, statistics, as well as cognitive and social psychology, neuroeconomics is defined as "an interdisciplinary field that seeks to explain human decision making, the ability to process multiple alternatives and to follow a course of action" [25]. Previous research endeavors in this field reported the existence of three main/universal types of e-Customers, regardless of the type of industry, including (1) 'Tightwads', (2) 'Average Spenders', and (3) 'Spendthrifts' [26]. Moreover, the latest research findings claim that in any population of e-Customers, 'Tightwads' comprise 24%, 'Average Spenders' cover 61% and 'Spendthrifts' involve 15% [27, 28]. Based on these three classes of e-Customers, a discrete random variable that resembles the operating profile, along with its probability mass function (pmf), can be defined. The operating profile defines the mix constituted by various e-Customer classes: if $k$ classes of e-Customers have been identified, ($t_1$, $t_2$, $t_3$ …, $t_k$), then each class can be associated a corresponding probability, drawn from the probability mass function vector ($p_1$, $p_2$, $p_3$ …, $p_k$), such that $\sum_{i=1}^{k} p_i = 1$. These probabilities are, in fact, a measure of the participation of each e-Customer class within the workload mix.

The second premise is related to the quantitative aspects of the workload characterization. The arrivals of e-Customers in an e-Commerce website can be mathematically modeled by a Poisson process, defined by the number of arrivals per time unit, i.e. the arrival rate $\lambda$ [e-Customers/s]. The times elapsing between each consecutive arrival comprise an i.i.d. (independent and identically distributed) random variable, exponentially distributed. Since the Markov property of the exponential distribution holds for each particular moment, the expected (mean) time to the next arrival is constant, given by $1/\lambda$. Moreover, let $\lambda$ be the total arrival intensity of e-Customers belonging to different classes ($t_1$, $t_2$, $t_3$ …, $t_k$), which comprise the workload mix. If the probability of classes' presence in the workload mix is represented by the probability vector ($p_1$, $p_2$, $p_3$ …, $p_k$), where $\sum_{i=1}^{k} p_i = 1$, then the arrival intensity of e-Customers, belonging to each particular class $t_i$ ($i = 1, 2, 3, …, k$), is given by the product $\lambda \times p_i$ ($i = 1, 2, 3, …, k$) [29].

## 6. Description of the Web-based Simulation Model

The Web-based simulation model has been built using Insight Maker®, and it entirely reflects the SD approach, especially Stock-and-Flow simulation. It is freely available for use at https://insightmaker.com/insight/34138/e-Commerce-Revenue-Estimator. The model provides a holistic insight into the e-Customers' dynamic behavior during online shopping sessions and combines both qualitative and quantitative aspects of the interaction between e-Customers and a hypothetic e-Commerce website.

Due to its robustness, the simulation model can be logically divided into four parts, which are going to be presented in more details as follows.

### 6.1.    Part #1: e-Customer Classes and the Operating Profile

The first part of the simulation model is depicted in Fig. 1. The flow entering the container 'New e-Shoppers' denotes the appearance of new potential e-Customers, been modeled as a Poisson process with adjustable intensity $\lambda$ [e-Customers/s]. The container 'New e-Shoppers' contains the total population of potential e-Customers, regardless on their online shopping behaviors. The adjustable variable 'Control1 %', being initially set to 24 [%], defines the portion of the total number of e-Customers that belong to the 'Tightwad' class. Another adjustable variable, 'Control2 %', separates the number of e-Customers that belong to two other classes, by initially setting the flow of 'Average Spender' e-Customers to 80.263 [%] (out of 76%), which yields exactly 61%. The rest of e-Customers (19.737 [%] out of 76%, which yields exactly 15%) flow into the container labeled 'Spendthrift e-Shoppers'.

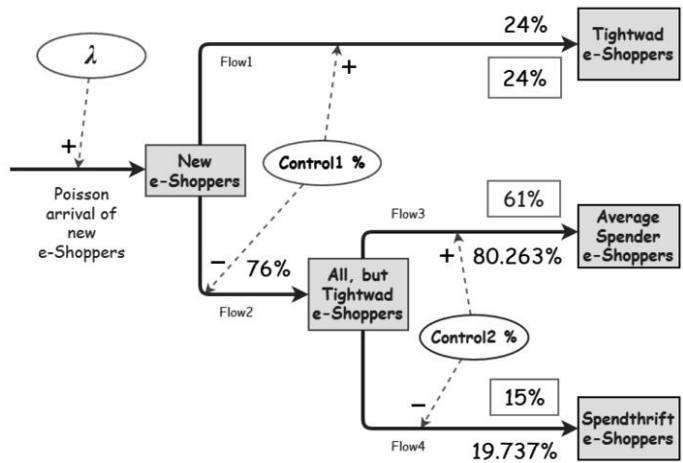

**Fig. 1.** E-Customer classes and the operating profile

In this manner, the first part of the simulation model introduces the three e-Customer classes as discussed in the previous section, i.e. $t_1$ = 'Tightwad', $t_2$ = 'Average Spender', and $t_3$ = 'Spendthrift' e-Customers. The vector of corresponding initial probabilities ($p_1$ = 0.24, $p_2$ = 0.61, $p_3$ = 0.15) defines the initial operating profile, i.e. the participation share of each particular e-Customer class into the workload mix. All simulation results described in this paper have been gained assuming these parameters' values. However, all of them, including many others, are easily adjustable online, so one can initiate multiple scenario runs just by moving a corresponding slider in the Web browser.

**6.2.    Part #2: Logic and Dynamics of Online Shopping Sessions**

Fig. 2 shows the second part of the simulation model: both the logic and the dynamics of 'Tightwad' e-Customers initiating an online session, which is identical, by its structure, for the two other classes of e-Customers.

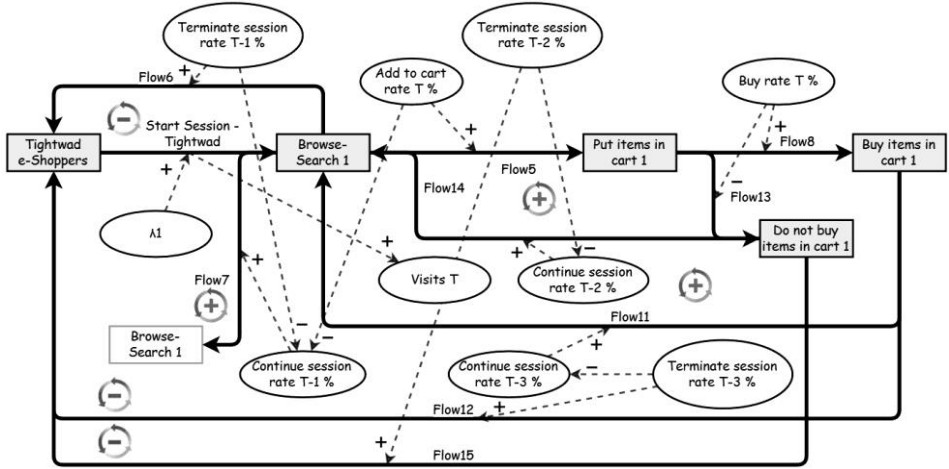

**Fig. 2.** Logic and dynamics of online shopping sessions
(a fragment that corresponds to 'Tightwad' e-Customers)

'Tightwad' e-Customers initiate online shopping sessions with an adjustable intensity $\lambda 1$. They can initially Browse/Search the website, and at any point in time they can either: (1) terminate the session; (2) continue browsing/searching, or (3) put item(s) into the shopping cart, all with corresponding rates. Those who will put the item(s) into the shopping cart can either buy them or not, leaving the shopping cart non-empty. In both cases, they can either (1) continue browsing/searching or (2) terminate their sessions, with corresponding rates.

**6.3.    Part #3: Estimating Total Sales Income**

The third part of the simulation model, which corresponds to the class of 'Tightwad' e-Customers, is presented in Fig. 3. In each particular time instance $t$, the container labeled 'Pay items in cart 1' contains the fraction of those e-Customers who have paid for the items put in the shopping cart. Knowing this number ($C_t$), and assuming that there are $M$ items in total available for selling, with buying probabilities (i.e. relative buying frequencies) $b_i$ ($i = 1, 2, …, M$) at selling prices $Pr_i$ ($i = 1, 2, …, M$), the revenue $R_t$, gained at time instance $t$, can be estimated by (2).

Based on (2), which is used for calculating the value of the output variable 'Revenue - Tightwad', one can estimate the cumulative revenue ($CR_T$), up to the time $T$, according to (3). Just for testing purposes, our simulation model includes only three items, whose

Evaluation of Business-Oriented Performance Metrics in e Commerce using Web-based Simulation 9

buying probabilities and selling prices are shown in Table 1. For the two other e-Customer classes, cumulative revenues are estimated in an identical manner. Fig. 4 portrays the fragment of the simulation model, needed to estimate the total cumulative revenue, given the cumulative values corresponding to each particular e-Customer class.

$$R_t = C_t \times \sum_{i=1}^{M} b_i \times Pr_i \qquad (2)$$

$$CR_T = \sum_{t=1}^{T} R_t \qquad (3)$$

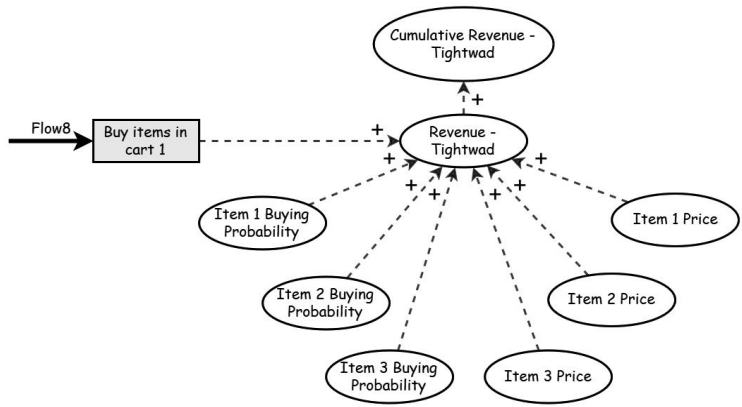

**Fig. 3.** Estimating total sales income ('Tightwad' e-Customers)

**Table 1.** Assumed buying probabilities $b_i$ ($i$ = 1, 2, …, $M$) and selling prices, $Pr_i$ ($i$ = 1, 2, …, $M$)

| Item # | Buying probability | Selling price |
|---|---|---|
| 1 | 0.3 | $6.00 |
| 2 | 0.1 | $10.00 |
| 3 | 0.6 | $2.00 |

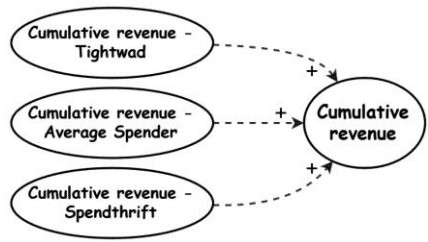

**Fig. 4.** Estimating the total sales income for all e-Customer classes

### 6.4.    Part #4: Estimating e-Commerce Business-Oriented Performance Metrics

Fig. 5 depicts the fragment of the simulation model that corresponds to the evaluation of all three business-oriented performance metrics, introduced previously.

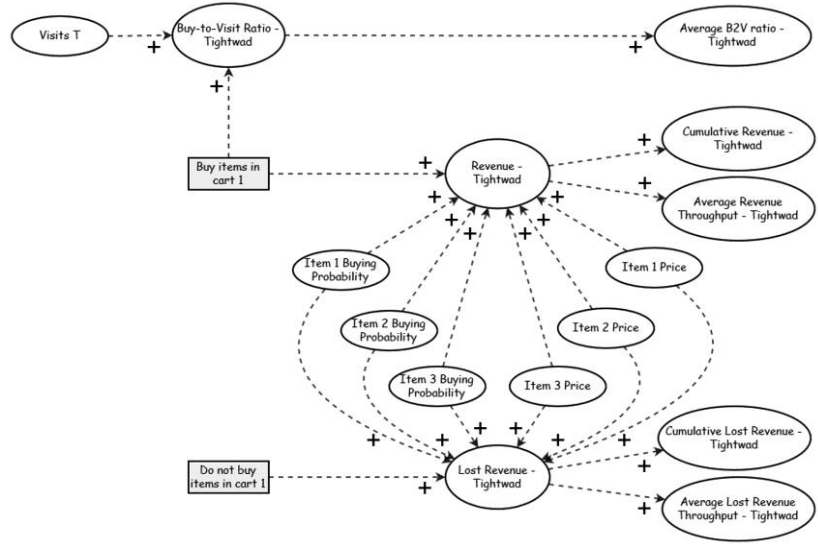

**Fig. 5.** Estimating the e-Commerce Business-Oriented Performance Metrics
(a fragment that corresponds to 'Tightwad e-Customers')

## 7.    Simulation Results

The simulation run took into account a time window of $T = 720$ [s], which is long enough for the simulation model to enter a steady state. It was accomplished using the parameter values shown in Table 2.

**Table 2.** Parameter settings

| Variable | Value |
|---|---|
| Control1 % | 24.000 (adjustable: [0.000, …, 100.000], step 0.001 ) |
| Control2 % | 80.263 (adjustable: [0.000, …, 100.000], step 0.001 ) |
| $\lambda$ | 1.1 (adjustable: [0.0, …, 50.0], step 0.1) |
| $\lambda 1$ | 5.5 (adjustable: [0.0, …, 50.0], step 0.1) |
| $\lambda 2$ | 3.5 (adjustable: [0.0, …, 50.0], step 0.1) |
| $\lambda 3$ | 1.5 (adjustable: [0.0, …, 50.0], step 0.1) |
| Buy rate T % | ~ N($\mu = 0.25$, $\sigma = 0.08333$) |
| Buy rate AS % | ~ N($\mu = 1.50$, $\sigma = 0.50000$) |
| Buy rate S % | ~ N($\mu = 5.00$, $\sigma = 1.66666$) |
| Add to cart rate T % | 5 (adjustable: [0.0, ..., 10.0], step 0.1) |
| Add to cart rate AS % | 20 (adjustable: [10.0, ..., 30.0], step 0.1) |
| Add to cart rate S % | 50 (adjustable: [30.0, ..., 70.0], step 0.1) |

Based on these assumptions, the dynamics of *estimated sales incomes* over time, for each particular e-Customer class, are graphically shown in Fig. 6, which is, also, a graphical representation of the *Revenue Throughput* [$/s] dynamics over time, for each particular e-Customer class.

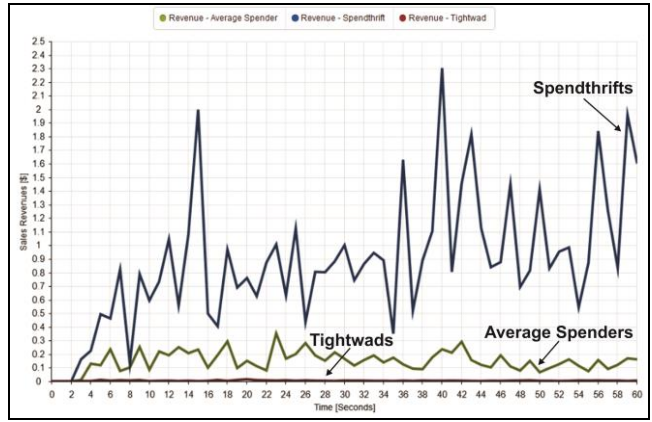

**Fig. 6.** The dynamics of estimated sales incomes over time ($T = 60$ [s]), which also represents the dynamics of Revenue Throughput [$/s] over time, for each particular e-Customer class

In addition, the dynamics of estimated *cumulative sales revenues* over time, for each particular e-Customer class, are graphically shown in Fig. 7. It is visually obvious that the trends for all three classes of e-Customers are approximately linear. The slope of each of them is, in fact, an exact measure of the revenue throughput intensity: the bigger the slope, the higher the Revenue Throughput [$/s]. The linearity trend approximation gives one an opportunity to analytically model and evaluate cumulative revenues [$] at a given point in time, as shown in Fig. 8.

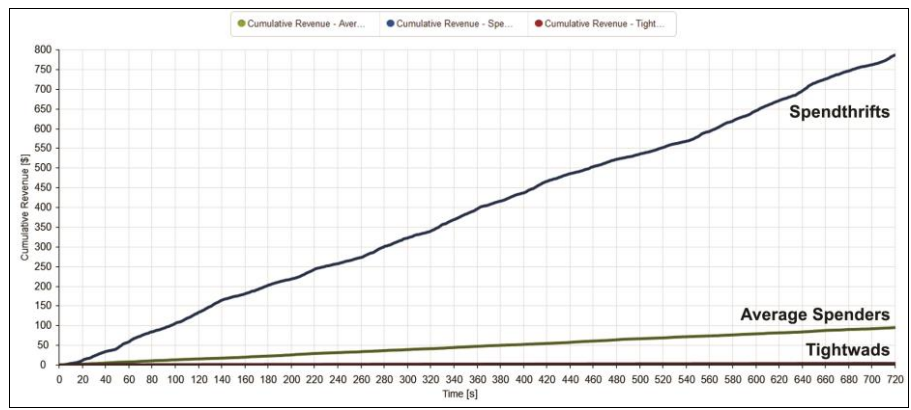

**Fig. 7.** The dynamics of estimated cumulative sales incomes over time ($T = 720$ [s]), for each particular e-Customer class

In all three cases, the high values of the coefficient of determination, $R^2$, indicate that a great percentage of variation can be explained by regression equations.

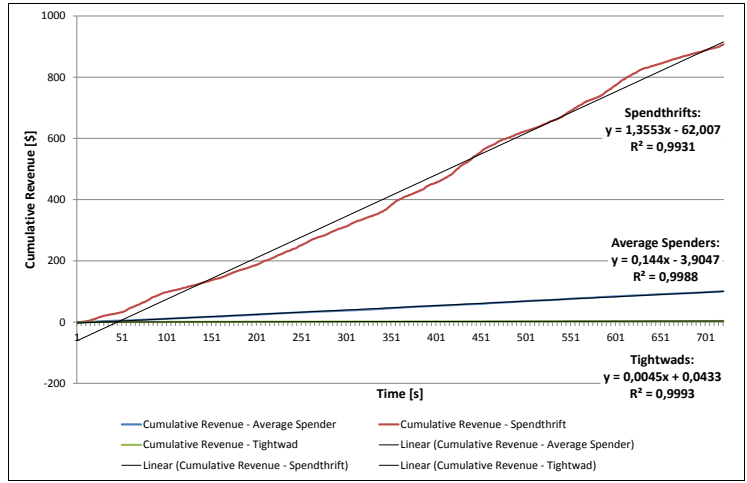

**Fig. 8.** A linear trend approximation of the cumulative revenues [$] over time ($T$ = 720 [s]), for all three classes of e-Customers

The estimated *total sales income* [$], which is shown in Fig. 9 is, simply, a sum of the three cumulative revenues trends, corresponding to the three e-Customer classes. Again, based on the simulation data, one can approximate the values with a linear trend, in order to analytically model and estimate the total sales income $I$ [$] at a specific point in time, as given by the linear regression equation (4).

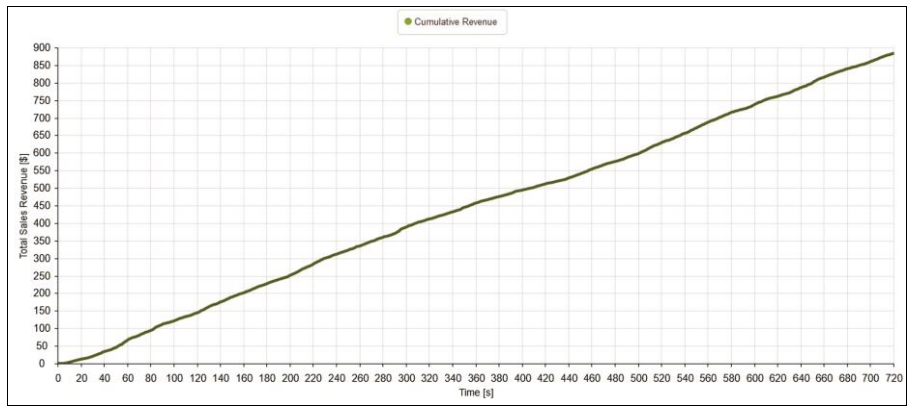

**Fig. 9.** The dynamics of estimated total sales income [$] over time ($T$ = 720 [s])

$$I[\$] = 1.2259 \times t[\text{s}] + 1.5048 \tag{4}$$

Evaluation of Business-Oriented Performance Metrics in e Commerce using Web-based Simulation
13

Regarding other e-Commerce business-specific performance metrics, the dynamics of *averaged Buy-to-Visit Ratio* [%] over time is depicted in Fig. 9. For 'Tightwad' e-Customers, the averaged Buy-to-Visit Ratio [%] is stabilizing within the interval [0.025, …, 0.028] [%], for 'Average Spenders' within the interval [1.24, …, 1.27] [%], and for 'Spendthrifts' within the interval [14.0, …, 14.5] [%].

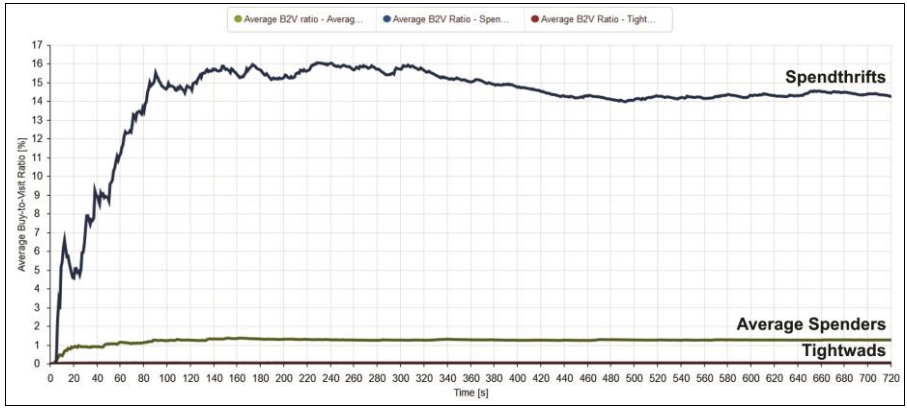

**Fig. 9.** Estimation of the averaged Buy-to-Visit Ratio [%] over time ($T = 720$ [s]), for each particular e-Customer class

In addition, the dynamics of the *averaged Revenue Throughput* [$/s] over time, for the three classes of e-Customers, is given on Fig. 10. The equilibrium is gained with the following values: for 'Tightwad' e-Customers within the interval [0.0045, …, 0.0048] [$/s], for 'Average Spenders' within the interval [0.12%, …, 0,15%] [$/s], and for 'Spendthrifts' within the interval [1.2, …, 1.3] [$/s].

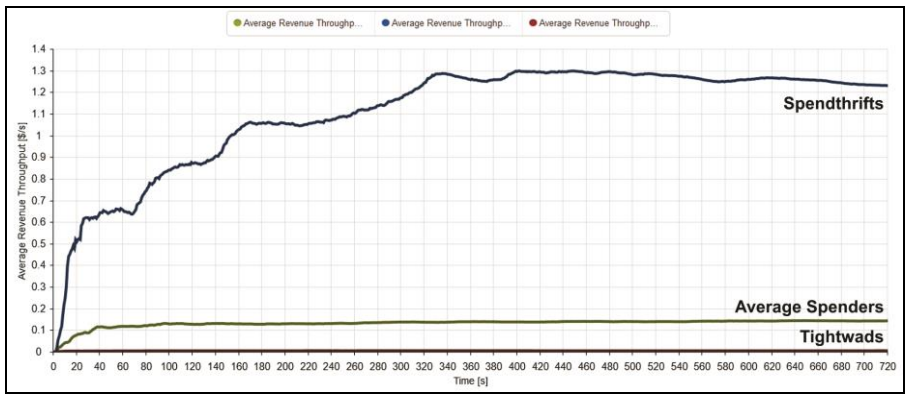

**Fig. 10.** Estimation of the averaged Revenue Throughput [$/s] over time ($T = 720$ [s]), for each particular e-Customer class

Finally, the dynamics of the *averaged Potential Loss Throughput* [$/s] is presented in Fig. 11. The highest values of this performance measure are evident with 'Spendthrift' e-Customers [20.0, …, 21.0] [$/s], while the class of 'Average Spenders' exhibits values within the range [8.70, …, 9.20] [$/s]. The class of 'Tightwads' produces values within the range [1.75, …, 1.85] [$/s].

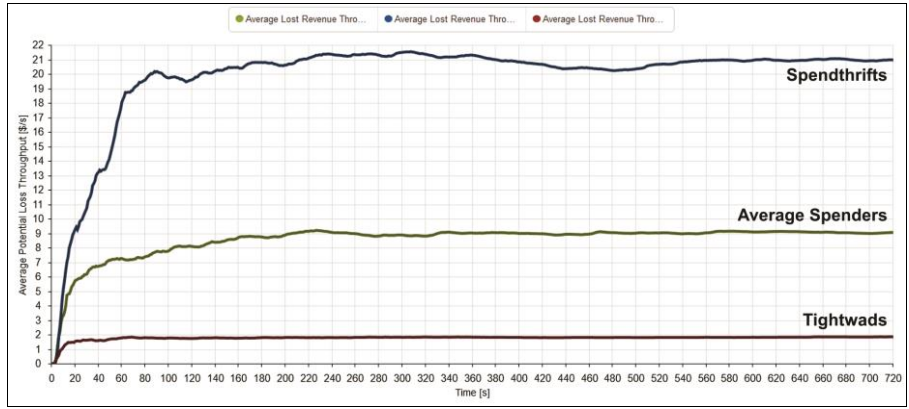

**Fig. 11.** Estimation of the averaged Potential Loss Throughput [$/s] over time ($T = 720$ [s]), for each particular e-Customer class

## 8.   Conclusion

In highly demanding online business environments, such as e-Commerce, estimating sales incomes is one of the most crucial tasks that can be successfully accomplished using computer simulations. The class of Web 2.0-based simulations, employing the system dynamics approach, can reveal new and significant insights into business processes, which, in turn, can increase their effectiveness, performances, and flexibility, thus creating an unprecedented competitive advantage for companies on a long run.

In the paper we have shown that, it is possible, under some real assumptions, to build a simulation model that, capturing e-Customer's online shopping behavior, is suitable for assessing not only the e-Commerce sales income dynamics, but also for evaluating other relevant business-oriented performance metrics, for various scenarios, each including a mixture of several e-Customer classes in various proportions. Such analyses and corresponding graphical presentations can be a valuable resource for e-Commerce websites' top management teams in creating their policies and in planning their operations on a strategic level.

In addition, Insight Maker® has proven to be a great innovative tool for mapping ideas by graphically visualizing them, and then, by converting maps into computational simulation models, to display specific behaviors and dynamics of the modeled system over time, as well as to carry out multiple scenario runs. Our simulation model is both updatable and upgradeable, and can be easily modified to include other performance measures. However, the main drawback of the system dynamics approach vis-à-vis our resulting simulation model could possibly be the fast increasing complexity of the model in the case when additional e-Customer classes and/or new items (and new selling prices) are introduced in the online store.

## References

1. Hinton, K., Chen, D.: The Fundamentals of Revenue Forecasting. pragmaticmarketing.com (–). [Online]. Available: http://pragmaticmarketing.com//resources/the-fundamentals-of-revenue-forecasting?p=1 (current January 2016)
2. Gifra, O.: e-Commerce Revenue Estimates. Oriol Gifra 360º: Sharing knowledge, opinions and thoughts (2011). [Online]. Available: http://www.oriolgifra.com/e-commerce-revenue-estimates/ (current January 2016)
3. Aganovic, Z.: Revenue Per Visit: The #1 metric to grow your e-Commerce revenue. HiConversion.com (2013). [Online]. Available: http://www.hiconversion.com/dollars-and-sense/revenue-per-visit-the-1-metric-to-grow-your-e-commerce-revenue/ (current January 2016)
4. Condra, C.: How to Estimate Site Traffic Based On Your Sales Goal. About Money/ Ecommerce @ About.com (2015). [Online]. Available: http://ecommerce.about.com/ od/eCommerce-Basics/ht/How-To-Estimate-Traffic-Needed-For-Sales-Goal.htm (current January 2016)
5. The Business Plan Shop Ltd.: How to forecast sales. thebusinessplanshop.com (–). [Online]. Available: https://www.thebusinessplanshop.com/blog/en/entry/how_to_forecast_sales (current January 2016)
6. Kearns, S.: How to Forecast Your Revenue. Intuit Inc. (2014). [Online]. Available: http://quickbooks.intuit.com/r/revenue/forecast-revenue (current January 2016)
7. Dugan, J.: Ecommerce Calculator for Measuring Revenue Growth. john-dugan.com (2014). [Online]. Available: https://john-dugan.com/ecommerce-calculator-for-measuring-revenue-growth/ (current January 2016)
8. Johnston, M.: The Website Revenue Guide: How Much Money Should Your Site Make? and How Much Money Should My Website Make?. MonetizePros.com (2013). [Online]. Available: http://monetizepros.com/monetization-basics/how-much-money-should-my-website-make-seven-real-world-website-revenue-statements/ and http://monetizepros.com/tools/how-much-money/ (current January 2016)
9. Menascé, D. A., Almeida, V. A. F.: Scaling for E-Business: Technologies, Models, Performance and Capacity Planning. Prentice Hall PTR, Upper Saddle River, New Jersey, NJ, USA. (2000)
10. Almeida, V. A. F.: Capacity Planning for Web Services: Techniques and Methodology. In: Calzarossa, M. C., Tucci, S. (eds.): Performance Evaluation of Complex Systems: Techniques and Tools. Lecture Notes in Computer Science (LNCS), Vol. 2459. Springer-Verlag, Berlin Heidelberg, 142–157. (2002)
11. Menascé, D. A., Almeida, V. A. F., Fonseca, R., Mendes, M. A.: Business-Oriented Resource Management Policies for e-Commerce Servers. Performance Evaluation – Special Issue on Internet Performance Modelling, Vol. 42, No. 2–3. Elsevier Science Publishers B. V., Amsterdam, The Netherlands. 223–239 (2000)
12. Harrell, C., Ghosh, B., Bowden, R.: Simulation Using ProModel, 2nd Edition. The McGraw-Hill Companies, Inc., New York, NY, USA. (2004)
13. Page, E. H., Griffin S. P., Rother, S. L.: Providing Conceptual Framework Support for Distributed Web-based Simulation within the Higher Level Architecture. In Proceedings of The SPIE Conference on Enabling Technologies for Simulation Science II, Orlando, FL, USA. 287–292. (1998)
14. Page, E., Opper, J. M.: Investigating the Application of Web-Based Simulation Principles within the Architecture for a Next-Generation Computer Generated Forces Model. Future Generation Computer Systems, Vol. 19, 159–169. (2000)
15. Byrne, J., Heavey, C., Byrne, P. J.: SIMCT: An Application of Web Based Simulation, In: Robinson, S., Taylor, S., Brailsford, S., Garnett, J. (eds.): Proceedings of The 2006

Operational Research Society (UK) 3rd Simulation Workshop (SW06). Royal Leamington Spa, UK. (2006)
16. Byrne, J., Heavey, C., Byrne, P. J.: A review of Web-based simulation and supporting tools, Simulation Modelling Practice and Theory, Vol. 18, No. 3, 253–276. (2010)
17. Harrell, C. R., Hicks, D. A.: Simulation Software Component Architecture for Simulation-based Enterprise Applications. In Proceedings of The 1998 Winter Simulation Conference (WSC '98). Washington, DC, USA. Vol. 2, 1717–1721. (1998)
18. Guru, A., Savory, P., Williams, R.: A Web-based Interface for Storing and Executing Simulation Models. In Proceedings of The 2000 Winter Simulation Conference (WSC '00). Orlando, FL, USA. Vol. 2, 1810–1814. (2000)
19. Pruyt, E.: Small System Dynamics Models for Big Issues: Triple Jump towards Real-World Dynamic Complexity, e-book, 1st ed., TU Delft Library, Delft, The Netherlands (2013). [Online]. Available: http://simulation.tbm.tudelft.nl/smallSDmodels/Intro.html (current January 2016)
20. Fortmann-Roe, S.: Insight Maker: A General-Purpose Tool for Web-based Modeling & Simulation. Simulation Modelling Practice and Theory, Vol. 47, 28–45. (2014)
21. Menascé, D. A., Almeida, V. A. F.: Capacity Planning for Web Services: Metrics, Models, and Methods. Prentice Hall PTR, Upper Saddle River, NJ, USA, 205–259. (2002)
22. Hristoski, I. S.: Performability Modeling and Evaluation of e-Commerce Systems. Ph.D. Dissertation. Faculty of Technical Sciences, "St. Clement of Ohrid" University, Bitola, Republic of Macedonia, 131–134. (2013)
23. Markellos, K., Markellou, P., Rigou, M., Sirmakessis, S.: Modeling the Behaviour of e-Customers. In Proceedings of The PCHCI 2001 Panhellenic Conference with International Participation in Human-Computer Interaction, Patras, Greece, 333–338. (2001)
24. Markellou, P., Rigou, M., Sirmakessis, S.: A Closer Look to the Online Consumer Behavior. In: Khosrow-Pour, M. (ed.): Encyclopedia of E-Commerce, E-Government and Mobile Commerce. Idea Group Publishing, Hershey, PA, USA, 106–111. (2006)
25. D-CIDES: Neuroeconomics. Duke Center for Interdisciplinary Decision Science, Duke University, Durham, NC, USA (2014). [Online]. Available: http://dibs.duke.edu/ research/d-cides/research/neuroeconomics (current January 2016)
26. Rick, S. I., Cryder, C., Loewestein G.: Tightwads and Spendthrifts. Social Science Research Network (SSRN) (2007). [Online]. Available: http://ssrn.com/ abstract=898080 (current January 2016)
27. Smith, J.: The 3 Types of Buyers, and How to Optimize for Each One. Neuromarketing by Roger Dooley (et al.) (–). [Online]. Available: http://www.neurosciencemarketing.com/ blog/articles/3-types-buyers.htm (current January 2016)
28. Patel, N.: How to Appeal to the Three Main Types of Buyers. The Daily Egg: Conversion Rate Optimization Made Easy (2015). [Online]. Available: http://blog.crazyegg.com/2015/ 01/06/3-types-of-buyers/ (current January 2016)
29. Stewart, W. J.: Probability, Markov Chains, Queues, and Simulation: The Mathematical Basis of Performance Modeling. Princeton University Press, New Jersey, NJ, USA, 385–394. (2009)